\documentclass{article}

\usepackage{PRIMEarxiv}

\usepackage[utf8]{inputenc} % allow utf-8 input
\usepackage[T1]{fontenc}    % use 8-bit T1 fonts
\usepackage{hyperref}       % hyperlinks
\usepackage{url}            % simple URL typesetting
\usepackage{booktabs}       % professional-quality tables
\usepackage{amsfonts}       % blackboard math symbols
\usepackage{nicefrac}       % compact symbols for 1/2, etc.
\usepackage{microtype}      % microtypography
\usepackage{lipsum}
\usepackage{fancyhdr}       % header
\usepackage{graphicx}       % graphics
\graphicspath{{media/}}     % organize your images and other figures under media/ folder
\usepackage{multirow}
\usepackage{xcolor}
\usepackage{soul}

\title{Star+: A New Multi-Domain Model for CTR Prediction
}

\author{
  Çağrı Yeşil\\
   \\
  Huawei Türkiye R\&D Center, İstanbul, Türkiye \\
  İstanbul\\
  \texttt{cagri.yesil1@huawei.com} \\
   \And
  Kaya Turgut \\
     \\
  Huawei Türkiye R\&D Center, İstanbul, Türkiye \\
  Eskişehir\\
  \texttt{kaya.turgut@huawei-partners.com} \\
}

\begin{document}
\maketitle

\begin{abstract}

In this paper, we introduce Star+, a novel multi-domain model for click-through rate (CTR) prediction inspired by the Star model. Traditional single-domain approaches and existing multi-task learning techniques face challenges in multi-domain environments due to their inability to capture domain-specific data distributions and complex inter-domain relationships. Star+ addresses these limitations by enhancing the interaction between shared and domain-specific information through various fusion strategies, such as add, adaptive add, concatenation, and gating fusions, to find the optimal balance between domain-specific and shared information. We also investigate the impact of different normalization techniques, including layer normalization, batch normalization, and partition normalization, on the performance of our model. Our extensive experiments on both industrial and public datasets demonstrate that Star+ significantly improves prediction accuracy and efficiency. This work contributes to the advancement of recommendation systems by providing a robust, scalable, and adaptive solution for multi-domain environments.

\end{abstract}

% keywords can be removed
\keywords{Recommendation \and CTR \and Multi-Domain}

\section{Introduction}

The explosive growth of online information and services has made recommendation systems essential for enhancing user experience and boosting platform revenue through personalized suggestions. Click-through rate (CTR) prediction, a critical component of these systems, has traditionally focused on single-domain environments, assuming homogeneity in user behavior and item interactions. However, large-scale commercial platforms operate across multiple business domains, each with distinct user behavior patterns and item characteristics. These multi-domain environments present unique challenges, including data sparsity, domain shifts, and the need for scalable models that can adapt to new domains.

Multi-domain CTR prediction aims to leverage shared knowledge across domains while capturing domain-specific characteristics. Traditional approaches often employ multi-task learning (MTL) techniques to project information into a common feature space \cite{ma2018modeling, tang2020progressive, zhou2023hinet}. However, these methods struggle to effectively model the complex relationships inherent among various domains and tasks, leading to suboptimal performance. Manual domain grouping based on business strategies is a common practice to reduce computational complexity, but it relies heavily on prior knowledge and may overlook underlying data distributions, thus limiting the model's representation capability.

Single-domain approaches in CTR prediction \cite{dilbaz2023stec, mao2023finalmlp, wang2021masknet, guo2017deepfm} are inadequate for multi-domain environments because they fail to capture domain-specific data distributions and cannot leverage the commonalities and distinctions between different domains. Also, using separate models for each domain is not preferable because it incurs high computational and maintenance costs due to the need for separate models for each domain. Also, the sparsity of data in low-traffic domains makes it impossible to train single-domain models on those domains.

To address these challenges, recent advancements \cite{li2022adaptdhm, yang2022adasparse, ning2023multi, zhou2023hinet, li2023adl, wang2022causalint, li2023hamur, sheng2021one, schoenauer2019multi} propose hierarchical and adaptive frameworks for specific to multi-domain problem. These approaches focus on intra-domain and inter-domain relations to make better prediction for each domain. Generally, these models proposes different solution to combine intra-domain and inter-domain information. The Star Topology Adaptive Recommender (Star) model \cite{sheng2021one} is one of the pioneer study in this domain where it combines intra-domain and inter-domain information with element-wise multiplication. 

In this paper, we proposed a new multi-domain model, Star+ that is mainly inspired from the Star model. With Star+, we aimed to investigate new interactions between shared and domain-specific parameters to optimize various metrics across multiple recommendation domains. We validated the effectiveness of our approach through extensive experiments on real-world industrial and public datasets, demonstrating significant improvements in prediction accuracy and efficiency. This work contributes to the advancement of recommendation systems by addressing the complexities of multi-domain environments and providing a robust solution for scalable and adaptive CTR prediction.

The main contributions of this work can be summarized as follows:

\begin{itemize}
  
  \item We propose Star+ along with different interaction strategies such as add, adaptive add, concatenate, and gating fusions to find the optimal balance between domain-specific and shared information.
  
  \item We evaluate layer normalization, batch normalization, and partition normalization on different datasets. Initially, partition normalization is proposed by the Star model for multi-domain and it is only tested on one private company dataset. By providing experiments on different datasets, we provide more information about these normalization techniques.

\end{itemize}

\section{Related Works}

This work is mainly inspired by advanced multi-domain learning research, especially the Star model \cite{sheng2021one}. Here, we mainly introduce the contents related to our research work.

The AdaSparse model \cite{yang2022adasparse} introduces domain-aware neuron-level weighting factors to measure the importance of neurons, and with that, for each domain, the model can prune redundant neurons to improve generalization. The EDDA model \cite{ning2023multi} uses graph-based networks to learn intra-domain and inter-domain information and combine them by using concatenation operation to make a final prediction.

In the CausalInt model\cite{wang2022causalint},  they introduce a new approach to avoid unnecessary information transfer from other domains to the target domain. For that, they use causal graph from the perspective of users and modeling processes, and then propose the Causal Inspired Intervention (CausalInt) framework for multi-domain recommendation. 

The AdaptDHM and ADL model \cite{li2022adaptdhm, li2023adl} focuses on distributions instead of domains. Therefore, it uses distribution-specific FCNs (Fully Connected Network) and shared FCNs. For that, it divides the dataset into K different distributions and decides to which distribution the given instance belongs with a distribution adaptation module. Unlike the Star model, it uses distribution-specific FCN and shared FCN separately and calculates the final output by multiplying the output of selected distribution FCN and shared FCN. Both models propose that it is enough to train $K+1$ MLPs where $K+1<M$, $K$ is number of cluster and $M$ is number of domains.

The HiNet model \cite{zhou2023hinet} attacks the multi-domain and multi-task problem at the same time. It uses a mixture-of-experts (MOE) component for each domain with a gating mechanism. Also, it defines another MOE for the shared network with a gating mechanism. They use a third component, a domain-aware attention network (SAN), to measure the importance of the information from other domains to the current one. Then, they concatenate the output of the three components and send it to the multi-task component.

The Hamur model \cite{li2023hamur} uses the adaptor logic utilized in large language models as a pluggable component to capture each domain’s unique characteristics. The parameters of this adapters are learnt with a domain-shared hyper-network which implicitly captures shared information among domains and dynamically generates the parameters for the adapter.

The Star model is composed of two main components: the star and the auxiliary FCN. In each layer $L$ of Star, there are domain weights $W_{d}^L$ for domain $d$ and shared weights $W_{s}^L$. While $W_{d}^L$ is only updated with the inputs from domain $d$, shared weight $W_{s}^L$ is updated with the inputs from all domains.

\begin{equation}
W^L = W_{d}^L \otimes W_{s}^L, b^L = b_{d}^L + b_{s}^L, 
\label{equation:star_weight}
\end{equation}

where $\otimes$ denotes the element-wise multiplication. Let $X_{d}^{L=0}$ denote the input from the d-th domain, The output $X_{d}^{L+1}$ is given by:

\begin{equation}
X_{d}^{L+1} = \phi(W^L X_{d}^L + b^L) 
\end{equation}

where $\phi$ is the activation function. The Auxiliary FCN is a simple FCN that uses all the data in all domains. In addition to Star FCN, it uses the domain indicator (domain ID) as a feature to distinguish the models. Denote the final output of star topology FCN as $s_s$ and the output of the auxiliary network as $s_a$. The final output is calculated with $sigmoid(s_a +s_s)$.

\section{The Proposed Approach}

In this section, we introduce the problem definition of multi-domain CTR prediction and then explain the technical details of our proposed Star+ approach.

\subsection{Problem Formulation}

The multi-domain CTR prediction aims to create a function $F: X \rightarrow Y$ that can accurately predict the label of a given advertisement among a set of business domains \{$D_1,..., D_M$\} where $M$ is the number of domains, $Y$ is label space, and $X$ is the common input feature space. Input feature space includes user information, user history, and item features that can be numerical, categorical, or sequential. The final form of $X$ is obtained by mapping all features to a low-dimensional vector space by using an embedding layer.

\subsection{Architecture Overview}

\begin{figure}
  \centering
  \includegraphics[width=0.5\textwidth]{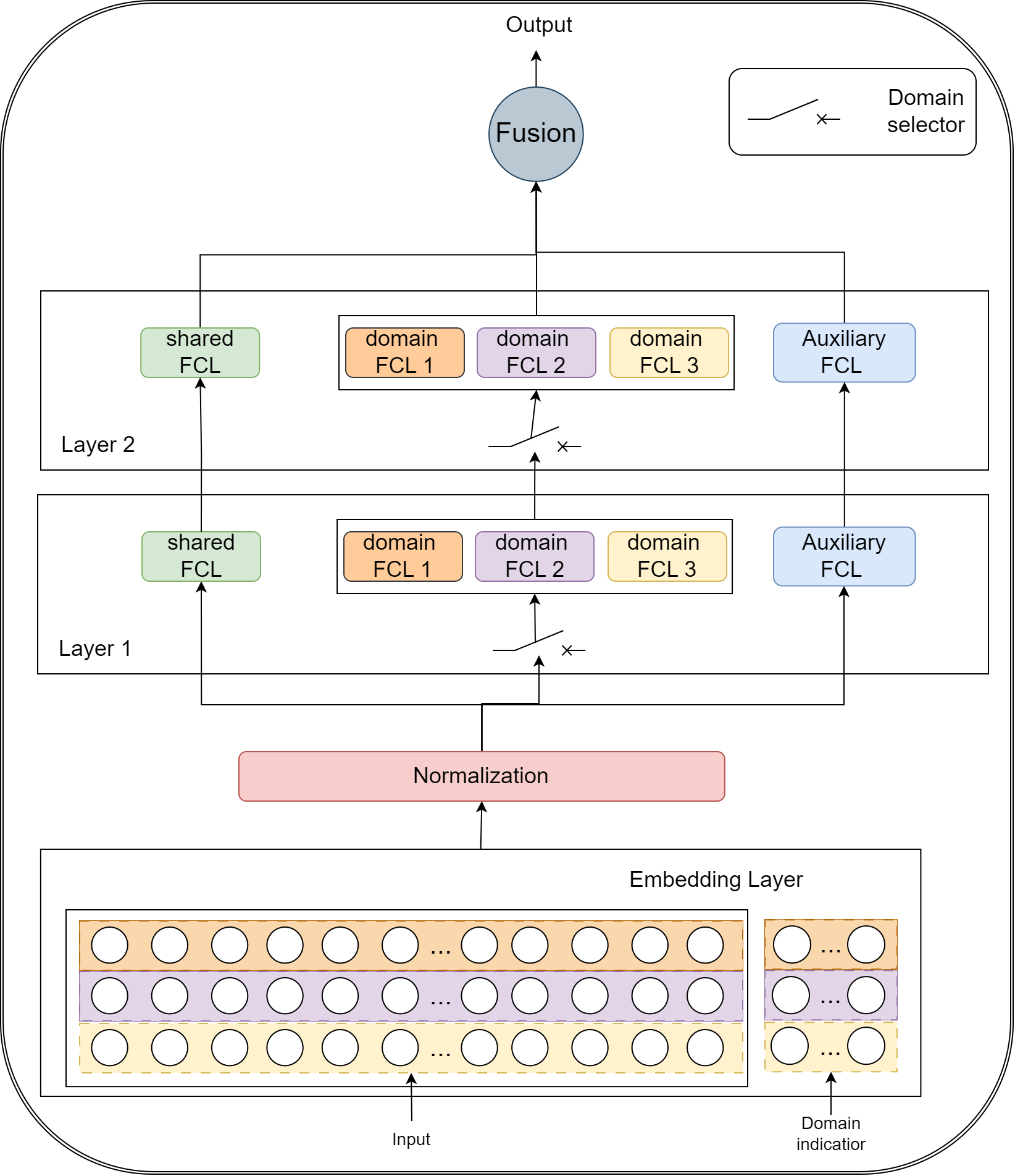}
  \caption{Architecture of Star+ model.}
  \label{fig:star_model}
\end{figure}

The Star+ model is mainly inspired by the Star model. There are three main components in the Star model: partition normalization, an auxiliary FCN, and the star component, which consists of a shared FCN and domain-specific FCNs. The partition normalization was first introduced in the Star paper. It is a variant of batch normalization that normalizes the input according to their domains. The star component aims to improve the prediction performance of a domain-specific FCN with the help of other domains. The shared FCN aims to learn common knowledge in all domains; therefore, it is updated with all domain data in the training process. However, a domain-specific FCN$_i$ focuses on knowledge specific to domain $i$. Therefore, in the training process, it is only updated with the data in domain $i$. The Star component combines the shared FCN and domain-specific FCN by multiplying the weights of the related FCN layers as given in \ref{equation:star_weight}. The auxiliary FCN is like the shared FCN; it uses all the data. But in addition to the shared FCN, it also uses domain indicators as a feature. Figure \ref{fig:star_model} shows the main architecture of the Star model composed of two layers and three domains for simplicity.

\begin{figure}
  \centering
  \includegraphics[width=0.5\textwidth]{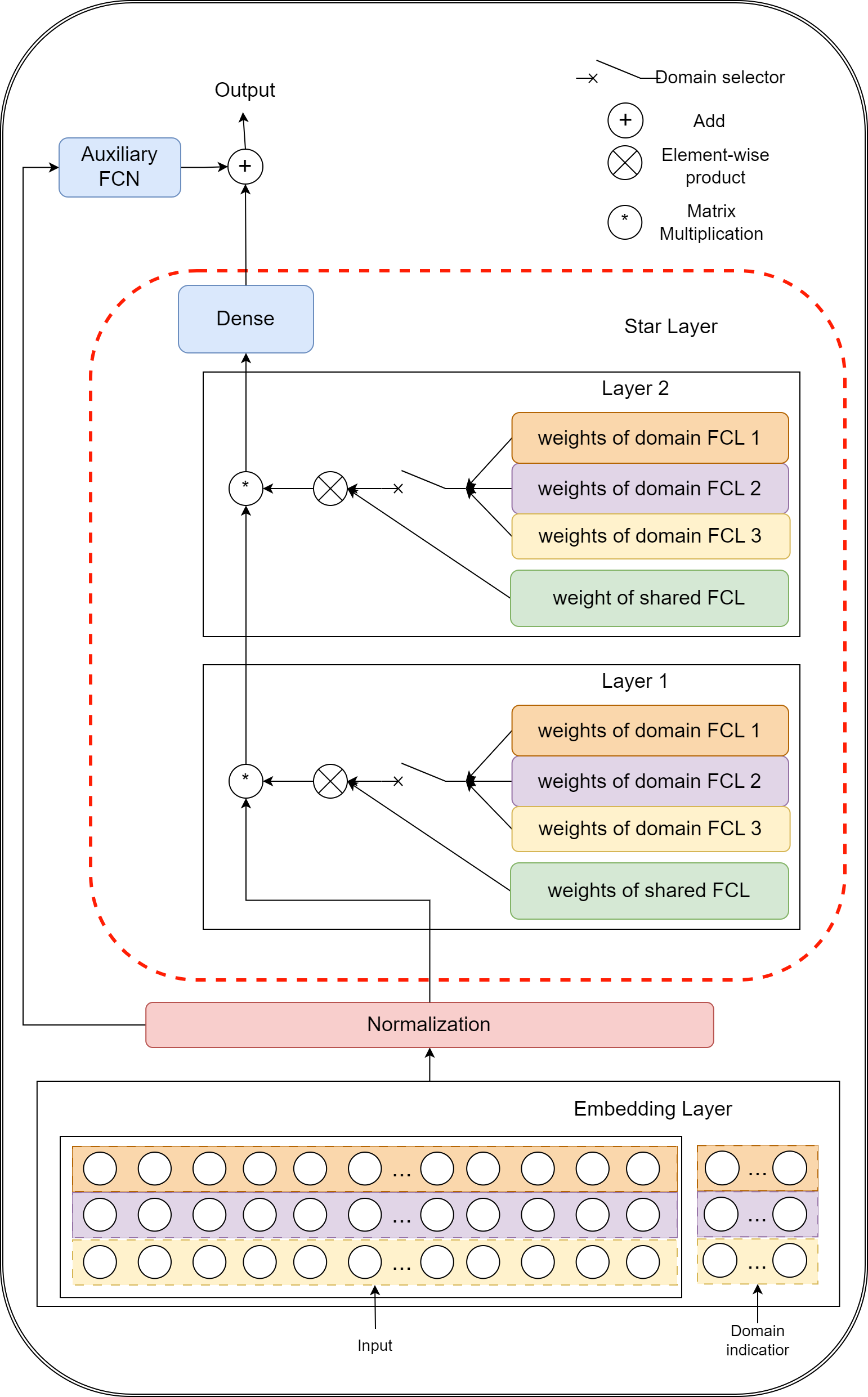}
  \caption{Architecture of Star model.}
  \label{fig:star}
\end{figure}

The Star model gives the same importance to domain and shared knowledge by directly multiplying the weights of shared and domain-specific layers. However, domain knowledge might be more vital for a domain with a large percentage of data while shared knowledge might be more vital for a domain with a lower percentage of data. 

In Star+, we use the same components as in Star, but this time we change how the shared and domain-specific layers interact with each other to cover the different relations of the domains with shared and domain-specific knowledge. For that, the Star+ model does not interact with the intermediate layers of shared and domain-specific FCN. It combines the final output of auxiliary, shared, and domain-specific FCNs with different fusion strategies. The fusion strategies are: add, adaptive add, concatenation, and gate, of which details are given in the next section. Simply put, the fusion strategies of the Star+ model aim to find a balance between domain knowledge and shared knowledge. Figure \ref{fig:star_model} shows the main architecture of the Star+ model, composed of two layers for simplicity. As seen in Figure \ref{fig:star_model}, the Star+ model uses normalization, too. In this study, we also investigated the effect of layer normalization, batch normalization, and partition normalization on the Star and Star+ models.

\subsection{Normalization Approaches}

As previously mentioned, the raw features are initially converted into low-dimensional embeddings to form the intermediate representation. This intermediate representation, denoted as $z$, is subjected to a normalization layer to ensure fast and stable training of deep networks. Batch normalization (BN) is a commonly used method among normalization techniques and has been shown to be essential for the successful training of deep neural networks. BN applies global normalization to all examples, accumulating normalization moments and learning shared parameters across the entire dataset. Specifically, the normalization process of BN during training is as follows:

\begin{equation}
z' = \gamma \frac{z-\mu}{\sqrt{\sigma^2 + \epsilon}} + \beta
\label{equation:batchnorm}
\end{equation}

where $z'$ is the output, $\gamma, \beta$ are learnable scale and bias parameters, and $\mu, \sigma^2$ are mean and variance of current mini-batch. On the other hand, Layer normalization (LayerNorm) is a technique used to stabilize and accelerate the training of deep neural networks by normalizing the inputs across the features of each training instance. Unlike batch normalization, which normalizes across the batch dimension, LayerNorm normalizes across the feature dimension, ensuring that the mean and variance are consistent within each training example. 

The Star paper proposed partition normalization (PN) to capture the unique data characteristics of each domain by using normalization statistics and parameters for different domains separately. Equation \ref{equation:partnorm} explains the partition normalization where $\gamma$, $\beta$ are the global scale and bias, and $\gamma_p$, $\beta_p$ are the domain-specific scale and bias parameters.

\begin{equation}
z' = (\gamma*\gamma_p) \frac{z-\mu}{\sqrt{\sigma^2 + \epsilon}} + (\beta+\beta_p)
\label{equation:partnorm}
\end{equation}

In this study, we conducted experiments by using all three normalization techniques on different datasets to measure the impact of each one.

\subsection{Star+ Topology FCN}

Similar to Star, Star+ uses different FCNs for different domains and one FCN for the shared domain. The interaction of domain FCNs and a shared FCN is the main contribution to this study. Also, it is the main difference from the Star model. In the Star+ model, the output of domain FCN and shared FCN is calculated independently. $X_{d}^{L+1}$ and $X_{s}^{L+1}$ are the outputs of layer $L$ in domain and shared FCNs and calculated as in Equation \ref{equation:domain_starp} and Equation \ref{equation:shared_starp}. $W_d^L$ and $W_s^L$ are domain and shared weights, $b_d^L$, and $b_s^L$ are domain and shared biases, and $X_{d}^L$ and $X^L$ are the input from domain d, and all domains respectively. 

In addition to domain and shared FCNs, we employed the auxiliary FCN used in the original Star paper for fair comparison. The auxiliary FCN is a simple FCN that uses all the data in all domains. The difference from shared FCN is that auxiliary FCN uses the domain indicator (domain ID) as a feature to distinguish the models.

\begin{equation}
X_{d}^{L+1} = \phi(W_d^L X_{d}^L + b_d^L) 
\label{equation:domain_starp}
\end{equation}

\begin{equation}
X_{s}^{L+1} = \phi(W_s^L X^L + b_s^L) 
\label{equation:shared_starp}
\end{equation}

The shared parameters are adjusted using the gradients from all examples, whereas the domain-specific parameters are updated only with examples from their respective domains. This approach enhances the model's ability to discern domain-specific differences for more precise CTR predictions while simultaneously learning commonalities across all domains through the shared parameters. Denote the final output of domain FCN, shared FCN and auxiliary FCN as $s_d \in R^o$, $s_s \in R^o$, and $s_a \in R^o$. The final output is calculated with Equation \ref{equation:final} where $\sigma$ is the sigmoid function and $f$ is the fusion function.

\begin{equation}
output = \sigma(f(s_d, s_s, s_a)) 
\label{equation:final}
\end{equation}

\subsection{Fusion strategies}

Generally, number of instances in each domain is significantly different. The domain with fewer instances has less data to learn patterns, therefore, it might be useful to give more weight to the shared and auxiliary networks. However, this strategy might be harmful for the domains that have large amount of data for a good learning process since the information that comes from the shared domains has a different distribution. Therefore, we proposed different fusion strategies that aims to balance the impact of shared and domain networks.

Denote $\hat{y_i}$ the output of the fusion function for the i-th instance and $y_i$ is the ground truth. Then, the aim is to minimize the cross entropy loss function between the $\hat{y_i}$ and $y_i$.

\begin{equation}
min - \sum_{n=1}^{N}  y_i log(\hat{y_i}) + (1-y_i)log(1-\hat{y_i})
\label{equation:loss}
\end{equation}

\subsubsection{Add}

In add strategies, we obtain the logits $s_d^{'}$, $s_s^{'}$ and $s_a^{'}$ from the $s_d$, $s_s$ and $s_a$. In other words, we apply a linear transformation to $s_d$, $s_s$, and $s_a$ for mapping $R^e \rightarrow R$ separately. Then, the final output is created by simply using a weighted sum of $s_d^{'}$, $s_s^{'}$ and $s_a^{'}$, as in Equation \ref{equation:weighted_sum}. The  $c_d$, $c_s$ and $c_a \in R$ are hyper-parameters which are constant during the training.

\begin{equation}
f(s_d, s_s, s_a) = c_d.s_d^{'} + c_s.s_s^{'} + c_s.s_a^{'}
\label{equation:weighted_sum}
\end{equation}

\subsubsection{Adaptive Add}

The adaptive add fusion aims to learn the weight of the domain and shared networks in the training process, unlike the add fusion. Therefore, it uses learnable parameters instead of hyper-parameters. It again uses Equation \ref{equation:weighted_sum}, but this time $c_d$, $c_d$ and $c_s$ parameters are defined as in Equation \ref{equation:adaptive_sum} where the $w_d\in R^{nd}$ are learnable weights where $nd$ is the number of domains and $\sigma$ is the sigmoid function. With the sigmoid function, the sum of $c_d$, $c_s$, and $c_a$ will be equal to 1. Since the shared and auxiliary networks use data from all the domains, we treated them as shared networks and distributed the weight equally.

\begin{equation}
c_d = \sigma(w_d),  c_s=c_a= \frac{1-\sigma(w_d)}{2}
\label{equation:adaptive_sum}
\end{equation}

\subsubsection{Gate}

\begin{figure}
  \centering
  \includegraphics[width=0.5\textwidth]{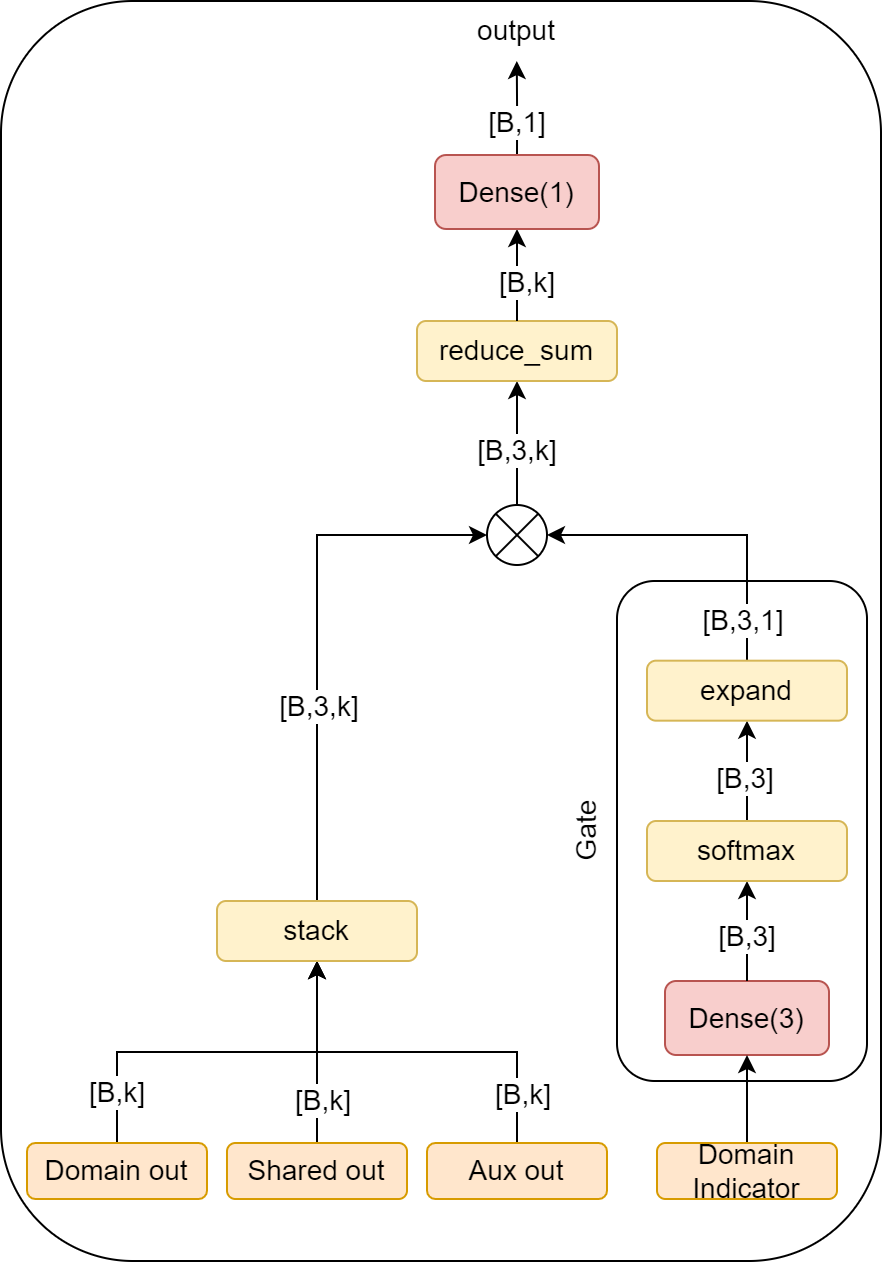}
  \caption{Architecture of gate fusion.}
  \label{fig:gate}
\end{figure}

Adaptive add fusion learns the importance of domain-specific and shared networks by using learnable weights for each domain. In the gate fusion strategy, the weights of each domain are learned by a gate network that uses the domain indicator feature as input. The gate network learns the weights of the domain, shared, and auxiliary networks for each instance separately, while adaptive add learns the global weight of each network where each instance uses the same. To be more specific, the flow of the gate fusion is illustrated in Figure \ref{fig:gate} where $B$ is the batch size, and $k$ is the output length of the domain and shared networks.

\subsubsection{Concatenation}

In this fusion strategy, we concatenate the $s_d$, $s_s$, and $s_a \in R^{nd}$ vectors into final vector $s \in R^{3*nd}$. Then, we apply a small FCN to $s$ to obtain the final logit. The number of layers in the FCN is a hyper-parameter. In concatenation fusion, the weights of the domain and shared networks are not explicitly learned since the outputs of those networks are concatenated. We aim to learn the domain relation implicitly by applying an FCN to the concatenated output.

\section{Experiments}

\subsection{Experimental Settings}
\textbf{Datasets.} We use two company datasets (company 1 and company 2) and one public dataset for the experiments. The training data of company datasets is collected from traffic logs of online display advertising system of Huawei. The data from seven days is used for training, and the data from the following day is used for validation and testing. Company datasets 1 and 2 consist of three and six domains, respectively. The Ali-CCP dataset \footnote{\url{https://tianchi.aliyun.com/dataset/408}}, sourced from Alimama’s traffic logs, is chosen and pre-processed following the protocol described in \cite{Xi2021AlicppPrep}. We utilize the contextual feature to segment the data into three distinct domains \cite{li2023hamur}.

Table \ref{tab:table_dataset} provides the data distribution percentage of each domain and average CTR value for each domain in the training set. It shows that different domains have different domain-specific data distributions, which can be reflected in the different CTRs. It can be seen that in Company 1, the domain with the lowest CTR (domain \#1) is 0.41\% while it constitutes 93.31\% of the whole dataset. In Company 2, the domain with the highest CTR (domain \#6) is 20.11\% while it constitutes 0.32\% of the whole dataset. Again, in Ali-CCP dataset, the domain with the lowest CTR (domain \#2) is 3.82\% while it constitutes 61.43\% of the whole dataset. Those major differences between domains create a challenge for recommendation models.

\textbf{Training.} To give a fair comparison, Star and Star+ models are trained with the same hyper-parameters. All models are trained with Adam [18], the learning rate is set to 0.001 and the batch size is 2000. We minimize the cross-entropy loss for samples from all domains to train the model.

\begin{table}
 \caption{The percentage and average CTR rate of each domain.}
  \centering
  \begin{tabular} {ccccccccc}
    \toprule
    & & all &1 &2 &3 &4 &5 &6\\
    
    \midrule
    \multirow{ 2}{*}{Company 1} & Percentage  & -         & 93.31\%  & 6.68\%   & 0.01\%   & - & -  & - \\
                                &    CTR      & 1.47\%    & 0.41\% & 16.28\% & 13.33\% & - & - & -    \\
    \midrule
    \multirow{ 2}{*}{Company 2} & Percentage  & -       & 59.76\%  & 16.09\%  & 15.59\%  & 6.28\% &  1.96\% &0.32\%\\
                                & CTR         &6.63\%   & 4.75\%   & 14.79\% & 2.94\% & 10.0\%  &13.4\% &20.11\%  \\
    \midrule
    \multirow{ 2}{*}{Ali-CCP}   & Percentage  & -       & 0.75\%   &61.43\%  & 37.82\%   & - & - & - \\
                                & CTR         & 3.9\%   & 4.4\%    &3.82\%   & 4.02\%    & - & - & -  \\ 
    
    \bottomrule
  \end{tabular}
  \label{tab:table_dataset}
\end{table}

\begin{table}
 \caption{Results of fusion techniques applied to company and public datasets.}
  \centering
  \begin{tabular}{lccccccccccc}
    \toprule
    Model     &Fusion Type & \multicolumn{2}{c}{Company 1}   & \multicolumn{2}{c}{Company 2} & \multicolumn{2}{c}{Public}\\
    \cmidrule(l){3-4} \cmidrule(l){5-6} \cmidrule(l){7-8}
    & &Loss &AUC &Loss &AUC  &Loss &AUC  \\
    \midrule
    Star  & -             & 0.15784           &0.93018          &0.64885        &0.81383          &\textbf{0.16218} &0.62067 \\
    Star+ & Add           &  0.15762           &0.93055           & \textbf{0.64417} & 0.81494          & 0.16232          &0.61968  \\
    Star+ & Concat        &  0.15787           &0.93056           & 0.64423          & 0.81542          & 0.16223          &0.62036  \\
    Star+ & Gate          &  \textbf{0.15738}  &0.93074           & 0.64439          & 0.81531          & 0.16239          &\textbf{0.62172}  \\
    Star+ & Adaptive Add  &  0.15745           & \textbf{0.93075} & 0.64509          & \textbf{0.81549} & 0.16254          &0.61972  \\
    \bottomrule
  \end{tabular}
  \label{tab:table_fusion_result}
\end{table}

\textbf{Metrics.} In evaluating the performance of a recommender system, particularly for CTR predictions in advertising, Area Under the Curve (AUC) and Logloss (i.e., logistic loss) are two crucial metrics. AUC, which stands for Area Under the Receiver Operating Characteristic (ROC) curve, assesses the model's capability to differentiate between positive (e.g., click) and negative (e.g., non-click) instances, particularly in terms of CTR. Since a higher AUC value signifies a higher true positive rate and a lower false positive rate, it indicates better model performance. It should be noted that for CTR prediction, a 0.1\% improvement is seen as a significant boost, as it can greatly enhance a company's revenue \cite{Shan2016DeepCross, Zhou2017DeepIN}. The second metric Logloss quantifies the accuracy of probabilistic predictions. A lower Logloss reflects the model's predicted probabilities align more closely with the actual outcomes.

\begin{table}
 \caption{Results of different normalization techniques on company and public datasets  in terms of AUC.}
  \centering
  \begin{tabular}{cccccccccc}
    \toprule
    Type     & \multicolumn{2}{c}{Company 1}   & \multicolumn{2}{c}{Company 2} & \multicolumn{2}{c}{Public}\\
    \cmidrule(l){2-3} \cmidrule(l){4-5} \cmidrule(l){6-7}
                    &   Star    &   Star+               &   Star    &   Star+            &Star &Star+ \\
    \midrule
    No Normalization&   0.93033 &   0.93041             &   0.81343 &   0.81453          &\textbf{0.62239}  &0.62216 \\
    LayerNorm       &   0.93056 &   \textbf{0.93075}    &   0.81508 &   \textbf{0.81549} &0.62158 &0.61972 \\
    BatchNorm       &   0.93029 &   0.93039             &   0.81370 &   0.81488          &0.62217 & 0.62042   \\
    PartitionNorm   &   0.93018 &   0.93032             &   0.81383 &   0.81468          &0.62067 &0.62085  \\
    \bottomrule
  \end{tabular}
  \label{tab:table_norm_result}
\end{table}

\subsection{Results}
To evaluate the performance of Star+ in predicting CTR for advertisements, we first assessed the impact of various fusion techniques, such as add, adaptive add, concatenation, and gate. Following this, we examined the effects of different normalization methods, including no-normalization (i.e., without normalization), layer normalization, batch normalization, and partition normalization. 

The experimental outcomes, including AUC and Logloss, are presented in Table \ref{tab:table_fusion_result}. In the initial experiment comparing fusion techniques, layer normalization was applied to all Star+ models, while partition normalization was used for Star due to its default setting. Overall, Star+ shows better performance compared to the Star model in company data across various fusion techniques, except for the add fusion method. Specifically, it achieves the highest AUC results with the adaptive addition fusion technique, recording improvements of 0.06\% and 0.204\% for the Company 1 and Company 2 datasets, respectively. Although it achieved high AUC scores, Star+ with adaptive addition showed higher loss values compared to other fusion techniques. It shows that it effectively distinguishes between positive and negative instances, but its probability predictions are less accurate. On the Ali-CPP dataset, Star+ with gate fusion achieved a 0.169\% increase in AUC, while Star recorded the lowest loss.

According to \cite{yilmaz2023correct}, different types of normalization yield varying responses across various datasets and models. Thus, to assess the effects of normalization techniques and ensure a fair comparison, various normalization methods were evaluated for both company and public datasets. The outcomes, based on AUC scores, are detailed in Table \ref{tab:table_norm_result}. For all normalization methods, the adaptive add fusion technique was used for Star+. For the company datasets, layer normalization provided the best AUC results for both Star and Star+. It appears that company datasets with similar structures in the feature space exhibit similar responses. Despite significant variations in data distributions across domains, Star+ consistently outperforms Star in company datasets. Notably, the non-normalized versions of Star and Star+ perform best for the Ali-CPP dataset, with Star showing better performance over Star+ with a 0.037\% increase. Conversely, partitioned normalization techniques in Star yield some of the lowest AUC results across all datasets, highlighting their limited effectiveness. It has become evident that the effectiveness of normalization techniques varies depending on the dataset.

\begin{table}
 \caption{AUC results per domain on company datasets.}
  \centering
  \begin{tabular} {clccccccc}
    \toprule
    & & all &1 &2 &3 &4 &5 &6\\
    
    \midrule
    \multirow{ 2}{*}{Company 1} & Star  & 0.93056 & 0.83279  & 0.68155   &  \textbf{0.54487}   & - & -  & - \\
                                & Star+      &  \textbf{0.93075}  &  \textbf{0.83315} 
                                &  \textbf{0.68388} & 0.44231 & - & - & -    \\
    \midrule
    \multirow{ 2}{*}{Company 2} & Star  & 0.81508 & 0.83815	& 0.64369	& 0.84288	& 0.66913	& 0.75271	& 0.64100\\
                                & Star+         &\textbf{0.81549}   & \textbf{0.83823}	& \textbf{0.64524} &\textbf{0.84353}	&\textbf{0.67086}	& \textbf{0.75418}	&\textbf{ 0.64442}  \\    
    \bottomrule
  \end{tabular}
  \label{tab:table_domain_auc}
\end{table}

In Table \ref{tab:table_domain_auc}, AUC results per domain for the company dataset are provided for Star and Star+ using layer normalization and the adaptive add fusion method. Star+ achieved the best value compared to the Star model in both datasets except for domain 3 on the Company 1 dataset. The data distribution of 0.01\% is believed to hinder both models from sufficiently learning the underlying information during the training phase, leading to increased randomness in their inferences. It has been demonstrated that Star+ effectively learns domain-specific characteristics for each domain.

\section{Conclusion}
In this paper, drawing inspiration from Star model, we propose the novel multi-domain Star+ approach as an enhanced method for predicting click-through rates (CTR). To capture interactions between shared and domain-specific features, Star+ employs various fusion strategies such as add, adaptive add, concatenation, and gating. Experimental evaluations on data from two company datasets and one public dataset showed that Star+ outperformed the Star approach in terms of AUC. Additionally, experiments were conducted using layer normalization, batch normalization, partition normalization, and no-normalization (i.e., without normalization) to observe their behaviors across different types of dataset. Layer normalization yielded better results for the company datasets, whereas normalization techniques were observed to have a negative impact on the public dataset.

%Bibliography
\bibliographystyle{unsrt}  
\bibliography{references}

\end{document}